\documentclass[aps,showpacs,superscriptaddress,nofootinbib,eqsecnum,prd,notitlepage,twocolumn]{revtex4-1} 

\pdfoutput=1

\usepackage{amsfonts}
\usepackage{amsmath}
\usepackage{amssymb}
\usepackage{graphicx,color}
\usepackage{float}
\usepackage{hyperref}
\usepackage{subfigure}
\usepackage{dcolumn}
\usepackage{soul}
\usepackage{ulem}
\usepackage{verbatim}
\usepackage{bm}  
\usepackage{slashed}

\begin{document}

\title{Dyson-Schwinger equation approach to Lorentz symmetry breaking with finite temperature and chemical potential}

\author{Y.M.P. Gomes}\email{yurimullergomes@gmail.com}
\affiliation{Instituto de Física - Universidade do Estado do Rio de Janeiro (UERJ), 20550-900, Rio de Janeiro, Rio de Janeiro, Brazil.}

\begin{abstract}
In this work, we investigate the dynamical breakdown of Lorentz symmetry in 4 dimensions by the condensation of a fermionic field described by a Dirac lagrangean with a four-fermion interaction. Using the Keldysh formalism we show that the Lorentz symmetry breaking modifies the Dyson-Schwinger equations of the fermionic propagator. We analyze the non-perturbative solutions for the Dyson-Schwinger equations using the combination of the rainbow and quenched approximations and show that, in equilibrium, the Lorentz symmetry breakdown can occur in the strong coupling regime and new features arise from this approach. Finally, we analyze the contributions of temperature and chemical potential and find the respective phase diagram of the model and analyze the dependence of the critical temperature and chemical potential as functions of the coupling constant.    
\end{abstract}

\maketitle
\section{Introduction}
Based on the fact that the standard model of particle physics (SM) is, despite its great success, unable to fully explain the phenomenology of particle physics, several models have been proposed to improve the knowledge about the quantum realm of fundamental particles, as supersymmetry, extra-dimensions, among others \cite{intro1,intro2,intro3}.

An attempt to extend the SM is to analyze extensions of the standard model with the introduction of Lorentz symmetry violation. In ref. \cite{string} the authors shown that in string theory could be possible that a spontaneous Lorentz symmetry breaking could occur and some phenomenological implications could arise in particle physics. This approach became a paradigm in Lorentz symmetry breaking. In particular, the simpler extension of the SM can be done by introducing new sectors where a vector field acquires a non-null expectation value in the vacuum. This background vector can be coupled with the standard model currents minimally or non-minimally and, in principle, could make contributions to the Charge-Parity (CP) symmetry violation or even to CPT symmetry violation \cite{lsv1,lsv2,lsv3,lsv4,lsv5,lsv6,lsv7,lsv8}.
 
Beyond the well-known works searching for LSV in particle physics, the approach of dynamical (instead of spontaneous) Lorentz symmetry breaking was implemented. 
In particular, a perturbative approach of dynamical Lorentz symmetry breaking (DLSB) is used to shed light on how this breakdown could occur in 4 and 3 dimensions \cite{DLSB1,DLSB2,DLSB3,
DLSB4,DLSB5,jenkins}. The main idea is that in analogy with the models of chiral symmetry breaking \cite{miranski}, the breakdown of the Lorentz symmetry appears from a fermion condensation through a four-fermion interaction. 

An example of this subject can be seen in ref. \cite{jenkins}, where a model with a vector four-fermion interaction $(\bar{\psi}\gamma_\mu \psi)^2$ is proposed and the dynamical symmetry breaking is analyzed. When the DLSB occurs, the vector acquires a non-null vacuum expectation value $\langle \bar{\psi} \gamma_\mu \psi \rangle \neq 0$, such way, if this DLSB generates a time-like constant vector, we can interpret this result as an effective chemical potential and its implications are well discussed in the paper.  

In the present work, we analyze the DLSB of an axial (or pseudo-vector) fermionic current starting from the following action proposed firstly by ref. \cite{DLSB1}:
\begin{equation}\label{eq1}
S = \int dx \left[ \bar{\psi} (i \gamma^\mu \partial_\mu - m) \psi -\frac{G}{2} (\bar{\psi} \gamma_\mu \gamma_5 \psi)^2 \right]~,
\end{equation}
where $G$ is the coupling constant and $G>0$ in order to maintain an attractive interaction and provide the fermionic condensation. We can introduce an auxiliary field $A_\mu$ such that the action can be rewritten as:
\begin{equation}
S = \int dx \left[ \bar{\psi} (i \gamma^\mu \partial_\mu - g \gamma^\mu A_\mu \gamma_5- m) \psi + \frac{g^2}{2 G} A_\mu A^\mu \right]~.
\end{equation}
In general, the $g$ parameter is introduced to expand the physical quantities and find quantum corrections. Although this is not our intention, we maintain $g$ explicitly to compare with the perturbative implementation. 

As can be seen in \cite{DLSB1, DLSB2}, in this perturbative approach the fermionic condensation can be calculated by a gap equation and it gives us the result  $\alpha_\mu = \langle A_\mu \rangle \propto G^{-1/2}$ showing that for a strong coupling constant $G$ we find a small value of DLSB parameter, and vice-versa. 
For instance, if $G \propto  M^{-2}$, with $M$ representing some mass scale, $\alpha_\mu \propto M$. As we will see in this paper, the non-perturbative approach gives us a very different result, and a phase transition will occur. Although we will discuss this result in our conclusions, we can already affirm that instead of a contribution for the chemical potential such can be seen in ref. \cite{jenkins}, our results can be interpreted as a chiral chemical potential $\mu_5$ above some critical coupling constant. The validity of calling $\mu_5$ a true chemical potential will be discussed in the conclusions. 

This work is organized as follows: In Section II we introduce the Keldysh formalism and through the generating functional we find the proper Dyson-Schwinger (DS) equations with the inclusion of Lorentz symmetry breaking. In Section III we focus on the DS equations of the full fermion propagator and show the approximations made. In Section IV we look for equilibrium solutions in some physical limits. In Section V we present our final comments and perspectives.

\section{Generating functional in Keldysh formalism}
In this work we adopt the real-time formalism in order to evaluate the contribution of the chemical potential and temperature. In this formalism, known as the Keldysh formalism \cite{kamenevbook}, we must double the fields ($\psi \rightarrow \{\psi^+,\psi^{-}\}$,$\bar{\psi} \rightarrow \{\bar{\psi}^+,\bar{\psi}^{-}\}$ , $A_\mu \rightarrow \{ A^{+},A^{-}\}$), and  the new action can be written as $S = S^+ - S^-$ and, applying the Keldysh rotations \cite{kamenevbook, kamenev}  
we can write the action $S[\bar{\psi},\psi,A] = S_0[\bar{\psi},\psi,A]+S_{int}[\bar{\psi},\psi,A]$, where:

\begin{eqnarray}\label{action1}\nonumber
S_0[\bar{\psi},\psi,A] &=& \int dx dy \Big[ \bar{\psi}^{a}(x) (S_0^{-1})_{ab}(x,y) \psi^{b}(y) + \\
&&\hspace{-1.5cm} +\frac{1}{2}  A_\mu^a(x)(D_0^{-1})_{ab}^{\mu \nu}(x,y) A_\mu^b(y) \Big] ,
\end{eqnarray}

\begin{equation}
S_{int}[\bar{\psi},\psi,A] = - g \int dx    \bar{\psi}^a(x)\gamma_{c,ab}^\mu \psi^b(x) A_{\mu}^c(x)
\end{equation}
with $(S_0^{-1})_{ab}(\partial) = (i \gamma^\mu \partial_\mu - m ) \hat{\sigma}^{cl}_{ab}$, $\gamma^\mu_{c,ab} = \frac{1}{\sqrt2}\gamma^\mu \gamma_5 (\hat{\sigma}_c)_{ab} $, $(D_0^{-1})^{\mu \nu}_{ab}(x,y) =\frac{g^2}{G} \delta(x-y) \hat{\sigma}^q_{ab} \eta^{\mu \nu}$, with $ \hat{\sigma}^{cl} = \begin{pmatrix}
1 & 0 \\ 0 & 1\\
\end{pmatrix}$ and $ \hat{\sigma}^{q}= \begin{pmatrix}
0 & 1 \\ 1 & 0 \\
\end{pmatrix}$, where $a, b, c = \{ cl, q\}$. We also have $A^{cl} = \frac{1}{\sqrt2}(A^+ + A ^-)$, $A^{q} = \frac{1}{\sqrt2}(A^+ - A ^-)$, $\psi^{cl} = \frac{1}{\sqrt2}(\psi^+ + \psi^-)$, $\psi^q = \frac{1}{\sqrt2}(\psi^+ - \psi^-)$,  $\bar{\psi}^{cl} = \frac{1}{\sqrt2}(\bar{\psi}^+ - \bar{\psi}^-)$ and  $\bar{\psi}^{q} = \frac{1}{\sqrt2}(\bar{\psi}^+ + \bar{\psi}^-)$. We emphasize that the index structure (with Latin indexes $a,b,c$) is due to the doubling of the fields in the Keldysh formalism, very similar to the closed-time-path (CTP) formalism \cite{CTP}, is necessary to construct the out-of-equilibrium QFT action.  A very instructive review of Keldysh formalism can be seen in ref. \cite{kamenev}.   
The generating functional is defined as follows \cite{Ryder}:

\begin{equation}
Z[\bar{\eta}, \eta, J] = \int D[\bar{\psi},\psi, A] e^{i S[\bar{\psi},\psi, A]}
\end{equation}
where $D[\bar{\psi},\psi, A] = \Pi_{a,b,c} d\bar{\psi}^a d\psi^b d A_\mu^c$. Since $A$ is not a gauge field, the necessity of the introduction of Faddeev-Popov ghosts is absent. We can define the generating functional of connected green functions as usual, i.e., $W = Ln Z$. We can write the quantum version of the field equations of motion (e.o.m.) of the fields as follows:
\begin{widetext}
\begin{eqnarray}\label{frstid}\nonumber
&&\int dy \Bigg \{  (D_0^{-1})_{\mu\nu}^{ab}(x,y) \frac{\delta W}{\delta J_\nu^b(y)} - g \delta(x-y) \frac{\delta W}{\delta \eta^b(x)}\gamma^\mu_{c,ab}\frac{\delta W}{\delta \bar{\eta}^c(y)} -g \delta(x-y) \frac{\delta } {\delta \eta^b(x)} \gamma^\mu_{c,ab} \frac{\delta W}{\delta \bar{\eta}^c(y)}\Bigg\}   + J_\mu^a(x) = 0\\
\end{eqnarray}
\end{widetext}

from the $A_\mu^a(x)$ e.o.m., and:
\begin{widetext}
\begin{equation}\label{secid}
\int dy \Bigg \{  (S_0^{-1})_{cd}(x,y) \frac{\delta W}{\delta \bar{\eta}^d(y)} - g \delta(x-y) \frac{\delta W}{\delta \bar{\eta}^d(x)}\gamma^\mu_{a,cd}\frac{\delta W}{\delta J^\mu_a(y)} -g \delta(x-y) \frac{\delta } {\delta \bar{\eta}^d(x)} \gamma^\mu_{a,cd} \frac{\delta W}{\delta J^\mu_a(y)}\Bigg\}   + \eta^c(x) = 0
\end{equation}
\end{widetext}
from $\bar{\psi}^c(x)$. Through \eqref{frstid} we can show an identity that will be important to our analysis that is the following:
\begin{eqnarray}\label{lsb}\nonumber
\frac{\delta W}{\delta J_\mu^{a}(y)}\Big{|}_{J=0} &=& \langle A^\mu_a(x) \rangle =\\\nonumber
&&\hspace{-2.cm} =-i g \int dz  Tr [\gamma_{\nu b,cd} S_{cd}(z,z)](D_0)^{\mu \nu}_{ab}  (z,x),\\
\end{eqnarray}
where $S_{bc}(x,y)= \frac{\delta^2 W}{\delta \eta^d(x)\delta\bar{\eta}^c(y)}\Big{|}_{J=0}$ the full fermion propagator and we use $J=0$ meaning $J_\mu^a = \bar{\eta}^a = \eta^a = 0$. We also use $Tr$ as the trace over the Dirac matrices space. Therefore, the $A_a^\mu$ field can assume a non-vanishing vacuum expectation value $\alpha_a^\mu(x) = \langle A^\mu_a(x) \rangle$ , i.e., a dynamical Lorentz symmetry breakdown can occur. Is important to highlight that in the Keldysh formalism the Lorentz symmetry breaking parameter acquire $x$-dependence.  
Going further, by the use of the effective action $\Gamma[\bar{\psi},\psi,A]$ defined as follows:
\begin{eqnarray}\nonumber
&&\Gamma [ \bar{\psi} , \psi , A ] = W [ \bar{\eta} , \eta , J ] + \\\nonumber
&&+\int dx \left [ \bar{\eta}^c (x) \psi_c (x) + \bar{\psi}^c(x) \eta_c (x) + A_\mu^a (x) J^\mu_a (x) \right ], \\
\end{eqnarray}
such way that the DS equation for the full bosonic propagator is given by 
\begin{equation}\label{dysonb}
(D^{-1})_{ab}^{\mu \nu} (x,y) - (D^{-1}_0)_{ab}^{\mu \nu} (x,y)  =\Pi_{ab}^{\mu \nu}(x,y),
\end{equation}
 where $(D^{-1})_{ab}^{\mu \nu}(x,y)= \frac{\delta^2 \Gamma}{\delta A^a_\mu(x)\delta A_\nu^b(y)}\Big{|}_{\Phi=0}$ is the inverse of the full bosonic propagator, $\Phi=0$  means $A_\mu^a=\alpha_\mu^a$ and $\bar{\psi}^a=\psi^a=0$. We also have that $\Pi_{ab}^{\mu \nu}(x,y)$, the polarization tensor, is given by:
 \begin{eqnarray}\nonumber
&&\Pi_{ab}^{\mu \nu}(x,y)= - i g \int dx' dy' \times\\\nonumber
&&Tr  \Big[\gamma^\mu_{a,cd}  S_{d d'}(x, y') \Gamma^\nu_{b,c' d'}(y,x',y') S_{c c'}(x,x') \Big],\\
 \end{eqnarray}
where $\Gamma^\nu_{b,c'd'}(y,x',y') = \frac{\delta^3 \Gamma}{\delta A_b^\nu(y) \delta \psi^{c'}(x')\delta\bar{\psi}^{d'}(y')}\Big{|}_{\Phi=0}$ is the full interaction vertex. The DS equation for the full fermionic propagator is the following:
\begin{eqnarray}\label{dysonf}\nonumber
S^{-1}_{cd}(x,y) - (S_0^{-1})_{cd}(x,y) &=& \\\nonumber
&&\hspace{-4.cm} = \Xi_{cd}(x,y) - g \delta(x-y)\gamma^\mu_{a,cd} \alpha_\mu^a(x), \\ 
\end{eqnarray}
where $S^{-1}_{cd}(x,y) = \frac{\delta^2 \Gamma}{\delta \psi^{c}(x)\delta \bar{\psi}^d(y)}\Big{|}_{\Phi=0}$, and the fermionic self-energy is given by:
\begin{eqnarray}\nonumber\label{self-energy}
\Xi_{cd}(x,y) &=& -i g \int dz dw \Big[\gamma^\mu_{a,cf} S_{fg}(x,z) \times \\
&&\Gamma^\nu_{b,dg}(w,y,z) D_{\mu \nu}^{ab}(x,w)\Big],
\end{eqnarray}
Finally, the $3$-point vertex respects the Bethe–Salpeter equation involving the 4-point vertex \cite{miranski} and so on.  As usual, we need to truncate the DS equations in some $n$-point vertex, and our choice will be clarified in the following section.

\section{Non-perturbative approach to DLSB}

Starting from the DS equation for the retarded component of the full fermion propagator:
\begin{eqnarray}\nonumber
S^{-1}_{cl cl}(x,y) &=& (S^R)^{-1}(x,y) = \\\nonumber
 &&\hspace{-2.cm} =(S_0^R)^{-1}(x,y) + \Xi^R(x,y) - g \delta(x-y)\gamma^\mu \gamma_5 \alpha_\mu (x), \\
\end{eqnarray}
where $\alpha_\mu = \alpha^{cl}_\mu$ (and $\alpha^q_\mu=0$). Now we need to choose the truncation of the vertex expansion, and we choose to use the so-called rainbow approximation, given by :
\begin{equation}
\Gamma^\nu_{b,dg}(w,y,z) \approx - g \gamma^\nu_{b,dg}\delta(w-y)\delta(z-y),
\end{equation}
which means that we are ignoring the contribution of the 4-point vertex. In QED this truncation violates the Ward-Takahashi identity of gauge invariance. But, due to the fact that our model does not have a gauge symmetry, this approximation good enough to our proposal. Therefore, from eq. \eqref{self-energy} the retarded component of the fermionic self-energy can be rewritten as follows:
\begin{eqnarray}\label{self-energy-R}\nonumber
\Xi^R(x,y) &=& i g^2  \gamma^\mu \gamma_5 S^K(x,y) \gamma^\nu \gamma_5  D_{\mu \nu}^{R}(y,x) +\\\nonumber
&&+i g^2 \gamma^\mu \gamma_5 S^R(x,y) \gamma^\nu \gamma_5  D_{\mu \nu}^{K}(x,y).\\
\end{eqnarray}
Assuming $D^R_{\mu \nu}(x,y) = \frac{G}{g^2}\eta_{\mu \nu} \delta(x-y) + O(g^2)$, $D_{\mu \nu}^K(x,y) =0+ O(g^2)$  (the so-called quenched approximation \cite{ref1}) and applying the Wigner transformation (see appendix I) we can rewrite the retarded component of the fermionic self-energy as:
\begin{equation}
\Xi^R(X,p) \approx i  G  \int dk \gamma^\mu \gamma_5 S^K(X,k+p)\gamma_\mu \gamma_5, 
\end{equation}
with $dk = \frac{d^4k}{(2 \pi)^4}$, $X = \frac{x+y}{2}$ and we assume a weak dependence in $X$-coordinates for all functions of the model, which means a small non-local character. Going further, we propose the following ansatz for the inverse of the full propagator:
\begin{equation}
(S^R)^{-1}(X,p) = A(X,p)\slashed{p}- \Sigma(X,p)- \slashed{B}(X,p)\gamma_5 ,
\end{equation}
where $A(X,p)$ is called renormalization function, $\Sigma(X,p)$ is the position-dependent mass function and $B_\mu(X,p)$ is introduced here in order to include the corrections to the parameter $\alpha_\mu$. Therefore, can be shown that the exact full fermion propagator can be written as follows \cite{fullprop}:

\begin{equation}
S^R = \frac{(A\slashed{p}- \slashed{B}\gamma_5 + \Sigma)(p^2 A^2 - \Sigma^2- B^2+ A[\slashed{p},\slashed{B}]\gamma_5)}{(p^2 A^2 - \Sigma^2- B^2)^2+ 4 A^2 ( p^2 B^2 - (p \cdot B)^2)+ i \epsilon},
\end{equation}

where we omit that $S^R,A,B$ and $\Sigma$ are functions of $X$ and $p$ for the sake of clarity. From the DS equation for the full fermionic propagator we reach:
\begin{equation}
(S^R)^{-1}(X,p) = \slashed{p}-m + \Xi^R(X,p) - g\gamma^\mu \gamma_5 \alpha_\mu (X),
\end{equation}
where 
\begin{eqnarray}\nonumber
 g \alpha_\mu(X) &=& - i  G ~ \int dp  Tr [\gamma_\mu \gamma_5 S^K(X,p)] = \\\nonumber
 &&\hspace{-1.cm} = 2 G Im \int dp Tr[\gamma_\mu \gamma_5 S^R(X,p) F(X,p)],\\
\end{eqnarray}
where we use that $S^K$ is the Keldysh component of the full propagator and is given by $S^K(X,p) = 2 i Im [S^R(X,p) F(X,p)]$ up to some $\partial_X$ contribution which we discard due to the small non-local approximation. The function $F(X,p)$ is called Wigner function (or Wigner quasiprobability distribution) and in our case is related to a generalization of the Fermi-Dirac distribution in out-of-equilibrium systems (see appendix II for details).

  With the appropriate traces over the Dirac matrices, we can find equations for the components of the full fermionic propagator and we can see that due to the quenched approximation the $p$-dependence of $B_\mu$ and $\Sigma$ disappears. Lastly, assuming that the fermionic Wigner function $F(X,p)$ is a real function and a scalar in the Dirac space and therefore $S^K(X,p)\approx 2 i Im [S^R(X,p)] F(X,p)$, we can write the equations for the components of the full fermion propagator as follows:

\begin{eqnarray}\nonumber
A(p,X) - 1  &=& -\frac{ G}{2p^2} \times\\\nonumber
 &&\hspace{-2.cm} Im\int dk \Big\{ Tr [\slashed{p}\gamma^\mu \gamma_5 S^R(X,k)\gamma_\mu \gamma_5]F(X,k) \Big\}, \\
\end{eqnarray}
\begin{eqnarray}\nonumber
B^\lambda(X) -  g \alpha^\lambda(X) &=&   -\frac{G}{2} \times \\\nonumber
&&\hspace{-3.cm} Im \int dk \Big\{Tr [\gamma^\lambda \gamma_5 \gamma^\mu \gamma_5 S^R(X,k)\gamma_\mu \gamma_5]F(X,k) \Big\},\\
\end{eqnarray}
and
\begin{eqnarray}\nonumber
\Sigma(X) - m =  \frac{G}{2}\times \\\nonumber 
&&\hspace{-3.cm} Im\int dk \Big\{ Tr [ \gamma^\mu \gamma_5 S^R(X,k)\gamma_\mu \gamma_5]F(X,k) \Big\},\\
\end{eqnarray}

\section{Equilibrium solutions:}
The results we achieve in the previous section are general and can be applied in a variety of physical contexts. The out-of-equilibrium QFT can be analyzed in a temperature gradient or chemical potential environment, or the presence of external fields and until in the cosmological context. The advantage of this approach is that it contains naturally the equilibrium as a solution and it will be our background from now on.

The equations which we find in the quenched and rainbow approximations are simplified since $A$, $B_\mu$ and $\Sigma$ are independent of $p$ and in a special case can also be local configurations (i.e., independent of $X$). This is already the case of equilibrium systems. Therefore, in this regime we find that:
\begin{widetext}
\begin{equation}
A(p)-1 = -\frac{4 G}{p^2} Im \int dk \frac{A(k) \left(k.p \left(A(k)^2 k^2-3 B^2-\Sigma ^2\right)+2 B.k B.p\right) }{(A(k)^2 k^2 - \Sigma^2- B^2)^2+ 4 A(k)^2 ( k^2 B^2 - (k \cdot B)^2)+ i \epsilon} F(k),
\end{equation}
\end{widetext}
where in equilibrium $F(k)=F_{eq}(k)= tanh\left( (k_0-\mu)/2T\right)$ (see appendix II). Since $A(p)-1 \propto \frac{1}{p}$ and using the fact that the integrand is an odd function of $k_\mu$ we will assume $A(p) = 1$ from now on. Using this assumption we reach: 
\begin{eqnarray}\label{gapt1}\nonumber
B_\mu &=& 4  G \times \\\nonumber
 &&\hspace{-1.cm}  Im \int dk \frac{ B_\mu \left(B^2-3 k^2+\Sigma^2\right)+ 2 k_\mu  (k \cdot B) }{D(k,\Sigma,B)+ i \epsilon}F(k),\\
\end{eqnarray}
\begin{equation}\label{gapt2}
\Sigma - m =  8 \Sigma   G Im \int dk \frac{ \left(B^2-k^2+\Sigma ^2\right)}{D(k,\Sigma,B)+ i \epsilon} F(k),
\end{equation}
where $D(k,\Sigma, B) = (k^2 - \Sigma^2- B^2)^2+ 4 ( k^2 B^2 - (k \cdot B)^2)$. Important to note that the DLSB parameter respects the identity $g \alpha_\mu = 2 B_\mu$. 
Going further, we study in the following subsections some special cases where the above equations have manageable solutions.
\subsection{$T=0$ and $\mu =0$:}
\begin{figure}[htb]
\centering
\includegraphics[width=1.\linewidth]{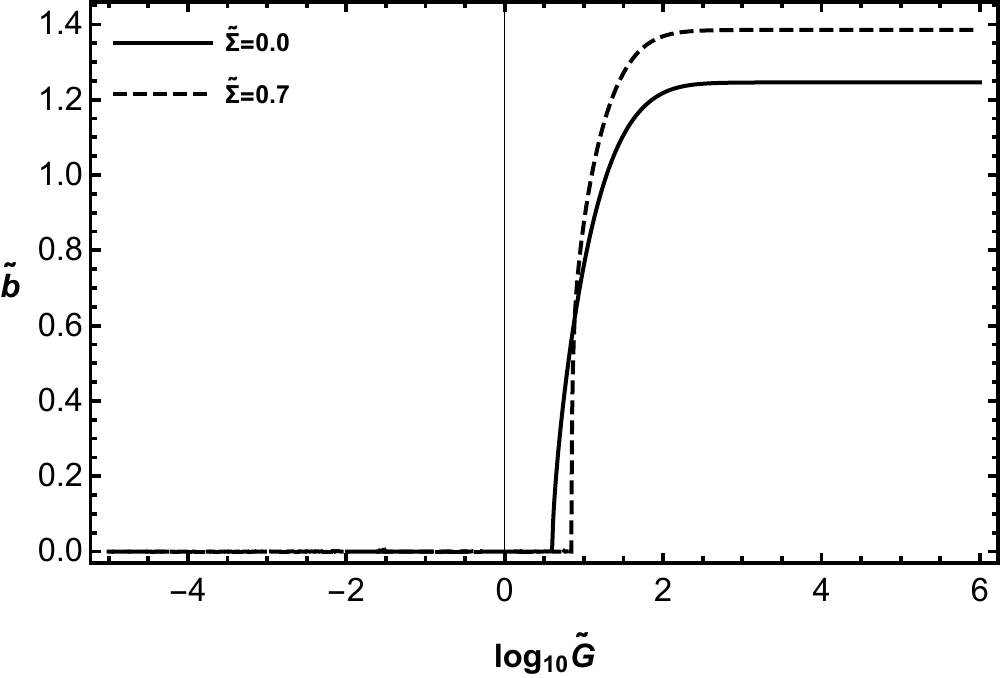}
\caption{Solution for $\tilde{b}$ in equation \eqref{gap1} for $\tilde{b}^2>0$; for $ \tilde{\Sigma} = 0$ (black curve) and for $\tilde{\Sigma}= 0.7$ (dashed curve), in units of $b$. }
\label{fig1}
\end{figure}

\begin{figure}[htb]
\centering
\includegraphics[width=1.\linewidth]{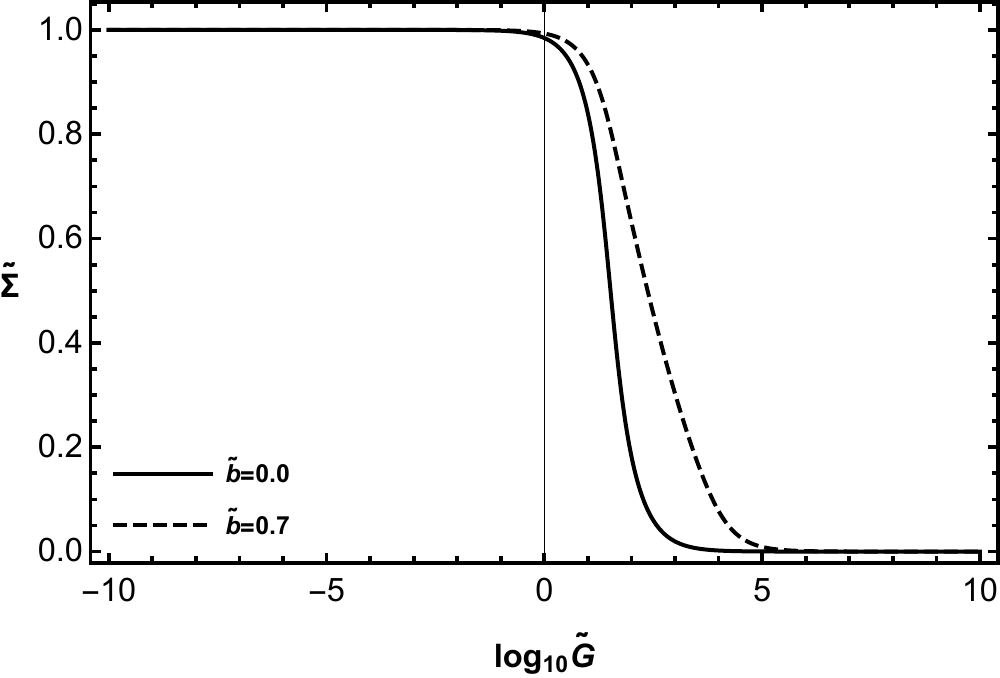}
\caption{Normalized solution of $\tilde{\Sigma}$ in equation \eqref{gap2} and $\tilde{b} = 0$ (black curve) and for $\tilde{b}= 0.7$ (dashed curve), in units of $m$. }
\label{fig2}
\end{figure}

In the limit $T,\mu \rightarrow 0$,  $F \rightarrow 1$ and is useful to apply the Euclidean rotation given by $k_0 = i k_4$, $B_0 = i B_4$ such that  $k^2 = - k_E^2$, $B^2 = - B_E^2$ and $k.B = - k_E . B_E$.  The gap equations can be rewritten as:

\begin{eqnarray}\nonumber
 &&(B_E)^2 +4 G  Re \int \frac{d^4k_E}{(2 \pi)^4}\times \\\nonumber
 &&~~ \Big[\frac{ ( B_E^2) \left(B_E^2-3 k_E^2-\Sigma^2\right)+ 2 (k_E \cdot B_E)^2 }{D(k_E,\Sigma,B_E)+ i \epsilon}\Big] = 0,\\
\end{eqnarray}
and 
\begin{eqnarray}\nonumber
&&\Sigma - m + 8 \Sigma   G Re \int \frac{d^4k_E}{(2 \pi)^4}\times \\\nonumber
&&~~~~~~\times\Big[ \frac{ \left(k_E^2 -B_E^2+\Sigma ^2\right)}{D(k_E,\Sigma,B_E)+ i \epsilon}\Big]=0,\\
\end{eqnarray}
where $D(k_E,\Sigma,B_E) = (k_E^2 + \Sigma^2- B_E^2)^2+ 4 ( k_E^2 B_E^2 - (k_E \cdot B_E)^2)$ and we contract eq. \eqref{gapt1} with $B_\mu$ after the Euclidean rotation. Using spherical coordinates in 4D Euclidean space as $k_1= |k_E| \cos \beta$, $k_2= |k_E|     \sin \beta \cos \theta $ , $k_3 = |k_E| \sin \beta \sin \theta \cos \phi$ and $k_4 = |k_E| \sin \beta \sin \theta \sin \phi$, fixing $B_E$ such that $B_E \cdot k_E = |B_E| |k_E| \cos \beta$ and integrating over $\beta \in [0,\pi]$, $\theta \in [0,\pi] $, $\phi \in [0, 2\pi]$,  $|k_E| \in [0,\Lambda]$ , where we introduce the cutoff parameter $\Lambda$, and $d^4k_E = d \theta \sin \theta d \phi d \beta \sin^2 \beta d |k_E||k_E|^3 $,  we find the following set of gap equations:

\begin{equation}\label{gap1}
\tilde{b}^2 + \tilde{G} F_1(\tilde{b},\tilde{\Sigma}) = 0 
\end{equation}
and
\begin{equation}\label{gap2}
 \tilde{\Sigma} - \tilde{m} + \tilde{G} \tilde{\Sigma} F_2(\tilde{b},\tilde{\Sigma}) =0
\end{equation} 
where $b = |B_E|$, and $\tilde{G} =  G \Lambda^2$, $\tilde{b} = |B_E|\Lambda^{-1}$ ,$\tilde{\Sigma} = \Sigma \Lambda^{-1}$, $\tilde{m} = m \Lambda^{-1}$ and omit the structure of $F_{1,2}$ for the sake of simplicity. In fig. \eqref{fig1} we can see the solution of \eqref{gap1} for some values of $\tilde{\Sigma}$ and in fig. \eqref{fig2} we can see the solution of \eqref{gap2} for some values of $\tilde{b}$.

\begin{figure}[htb]
\centering
\includegraphics[width=1.\linewidth]{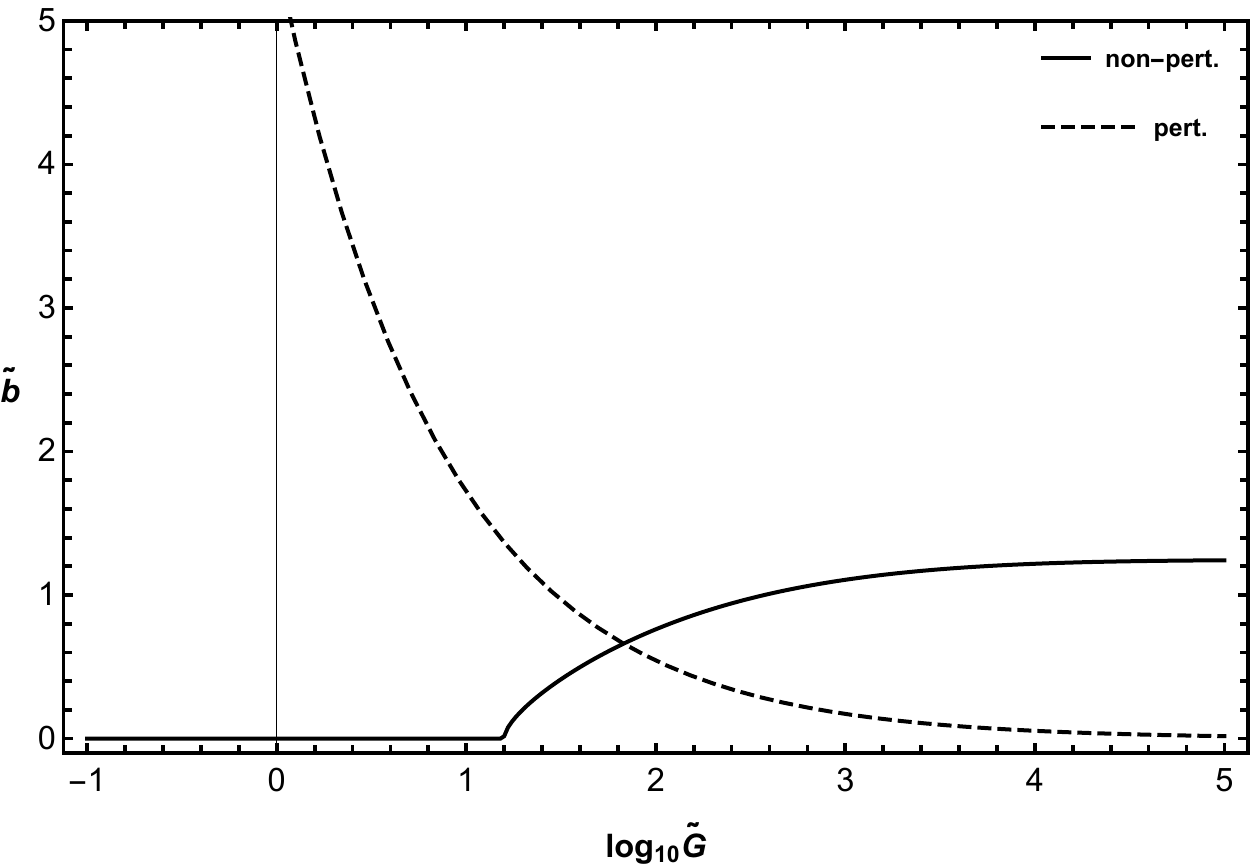}
\caption{Plot of the gap equation solution for $m=0$ given by eq. \eqref{gapmassless} (black curve) and the perturbative solution $\tilde{b} = \sqrt{3 \pi^2/G}$  \cite{DLSB1,DLSB2} (dashed curve). }
\label{fig3}
\end{figure}

If we assume $\tilde{b} =0$, eq. \eqref{gap2} reduces to $\tilde{\Sigma}-\tilde{m} +\frac{G}{2 \pi ^2} \tilde{\Sigma}  \left(\tilde{\Sigma} ^2 \ln \left(\frac{\tilde{\Sigma} ^2}{\tilde{\Sigma} ^2+1}\right)+1\right)=0$, and it's solutions is shown by the black curve of figure \eqref{fig2}. In the case of massless fermions ($m=0$), the solution $\tilde{\Sigma} = 0$ is possible and in this case we find the following  gap equation for $\tilde{b}$ :

\begin{eqnarray}\label{gapmassless}\nonumber
\frac{1}{ \tilde{G}} -\frac{ \left(15+\tilde{b}^2\left(48  \ln2 -31 + 24  \ln \left(\frac{\tilde{b}^2}{\tilde{b}^2+1}\right)\right)\right)}{24 \pi ^2}=0~.\\
\end{eqnarray}
As we can see in fig. \eqref{fig3}, the solution is very different of the (massless) perturbative result \cite{DLSB1,DLSB2} given by $b = \sqrt{\frac{3 \pi^2}{|G|}}$, which goes to infinity where $G$ approaches to zero, and goes to zero in the limit $G \rightarrow \infty$. Our result has (in the massless case) a critical coupling  $\tilde{G}_c = \frac{24 \pi^2}{15} \approx 15.8$, and for values of $\tilde{G}> \tilde{G}_c$ the   Lorentz symmetry breaking occurs. Is important to highlight that we only find time-like solutions, i.e., solutions where $\tilde{b}^2>0$.  

\begin{figure}[htb]
\centering
\includegraphics[width=1.\linewidth]{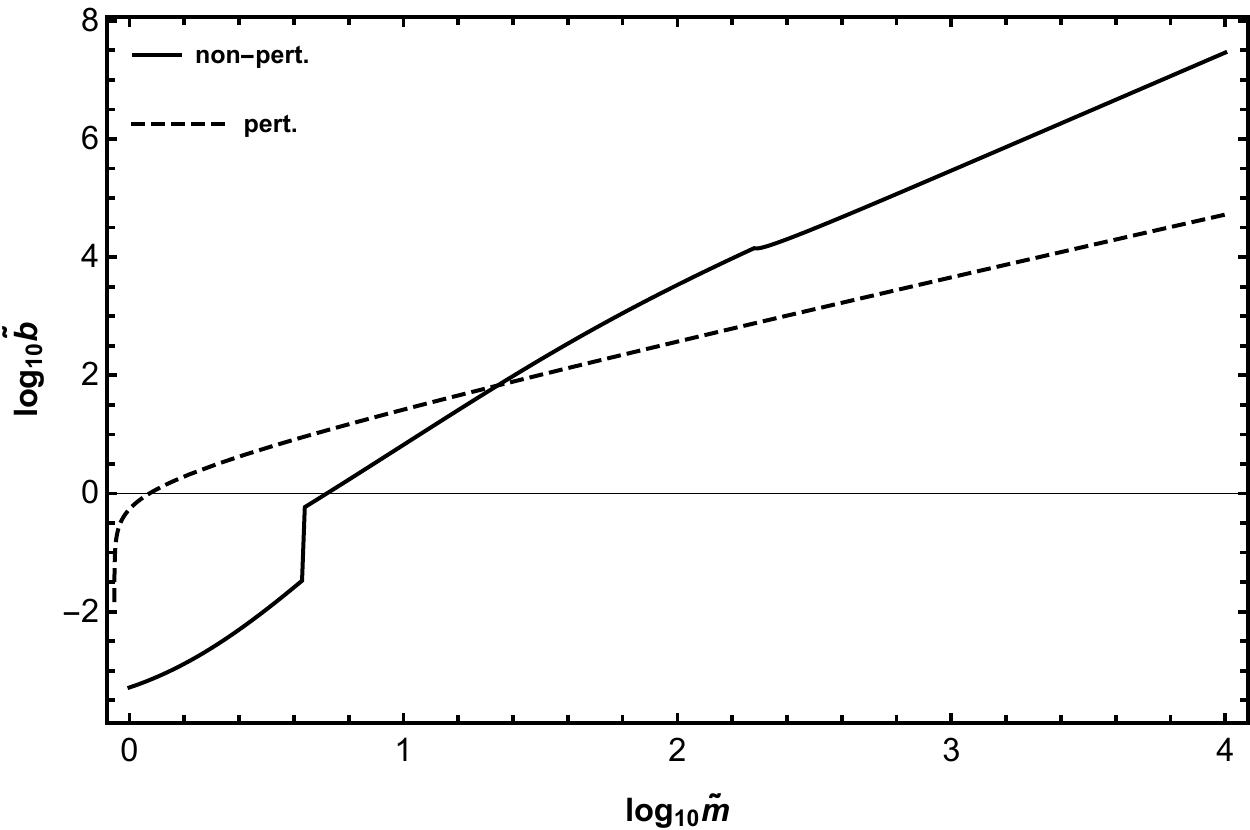}
\caption{Plot of the gap equation solution $
 \tilde{b}^2+\tilde{G} F_1(\tilde{b},\tilde{\Sigma})=0 $ for $\Sigma \approx  m$  and for $\log_{10} \tilde{G} = -2$ (black), and the perturbative solution for $\log_{10} \tilde{G} = -2$ , normalized in units of $\mu$ (black and dashed). }
\label{fig4}
\end{figure}

We can improve our result for $\tilde{b}$ assuming that $\Sigma \approx m$. This approximation is valid for $\tilde{G}<1$. In this approximation the gap equation is given by $
 \tilde{b}^2+\tilde{G} F_1(\tilde{b},\tilde{m})=0 $ and we can see in fig. \eqref{fig4} the plot of  $\tilde{b}$ in function of $\tilde{m}$ for $\tilde{G}= 10^{-2}$ and the perturbative result $b^2 = 3 \pi^2 \left( \frac{1}{|G|} + \frac{m^2}{2 \pi^2} \ln \left( \frac{m^2}{\mu^2}\right) \right)$ from ref. \cite{DLSB1}. In the limit $\tilde{m} \rightarrow 0$ the DLSB parameter $b$ vanishes, whereas the perturbative result is given by $b^2 = \frac{3 \pi^2}{|G|}$.    
 
\subsection{$T=0$ and $\mu \neq 0$:}
As we have shown in the previous section, the vector $B_\mu$ can acquire a non-null vacuum expectation value $\langle B \rangle = b$ at zero temperature and chemical potential, even in the non-perturbative regime.  In the case where we consider the global U(1) symmetry of the action in eq. \eqref{eq1}, a chemical potential must be added and this is achieved applying the shift in the equilibrium fermionic function $F(k_0) \rightarrow F(k_0 - \mu)$ where $\mu$ is the chemical potential. Explicitly, the fermionic function will be given as follows:
\begin{equation}
F(k_0 - \mu) = 1 +  \left( \frac{1}{e^{\beta(k_0-\mu)}+1}+  \frac{1}{e^{\beta(k_0+\mu)}+1} \right)\Theta(k_0)
\end{equation}  
where $\beta = T^{-1}$ and $\Theta(x)$ is defined as $\Theta(x) = 1$ for $x>0$ and $\Theta(x)= 0$ for $x<0$. In the regime where $T$ is negligible ($T<<\mu$) we can use the Sommerfeld expansion as follows. Let $H$ a function of $k_0$, therefore:

\begin{eqnarray}\label{sommer}\nonumber
I(\mu) &=& \int_{-\infty}^{\infty} dk_0 H(k_0)F(k_0-\mu)  \\
&=&I(0) + 2 \int_{0}^{\mu} dk_0 H(k_0).  
\end{eqnarray}
Therefore, the gap equation for $B$ can be rewritten as follows:

\begin{equation}
B^2(\mu) = b^2 - 8 \pi g G \left(   2 \int_{0}^{\mu} dk_0 H(k_0) \right) ,
\end{equation}
where 
\begin{eqnarray}\nonumber
H(k_0) &=&
\int \frac{d^3\vec{k}}{(2 \pi)^4} \Big[\\\nonumber
&&\hspace{-1.cm} B^2 \left(B^2-3 k^2+\Sigma^2\right)+ 2 (k \cdot B)^2 \Big] \delta(D(k)),\\
\end{eqnarray}
with $D(k) = (k^2 - \Sigma^2- B^2)^2+ 4 ( k^2 B^2 - (k \cdot B)^2)$ \footnote{We use the identity $Im \frac{1}{D(k) + i \epsilon} = \pi \delta(D(k))$}. In the case of time-like DLSB parameter we can choose a reference frame such that  $B_\nu = \delta_{\nu 0}B$  (also $b_\nu = \delta_{\nu 0} b$ ) , and in the massless case we can assume $\Sigma=0$,
and using the minimal subtraction scheme (MS) in the divergent component  $B(\mu=0) = b + (\text{divergent terms}) $, with $b$ finite, we find:
\begin{equation}
B^2 - b^2 + \frac{B G \mu \left(15 B^2 +\mu^2\right)}{6 \pi ^2}  = 0 \text{~~, if~~} B>\mu ,
\end{equation}
\begin{eqnarray}\nonumber
&&B^2 -b^2 +  \frac{ B^2 G \left(3 B^2 \ln \left(\frac{\mu }{B}\right)+5 B^2+3 \mu ^2\right)}{3 \pi ^2}  = 0 ,\\
&&\text{~~if~~} B<\mu
\end{eqnarray}

\begin{figure}[htb]
\centering

\includegraphics[width=1.\linewidth]{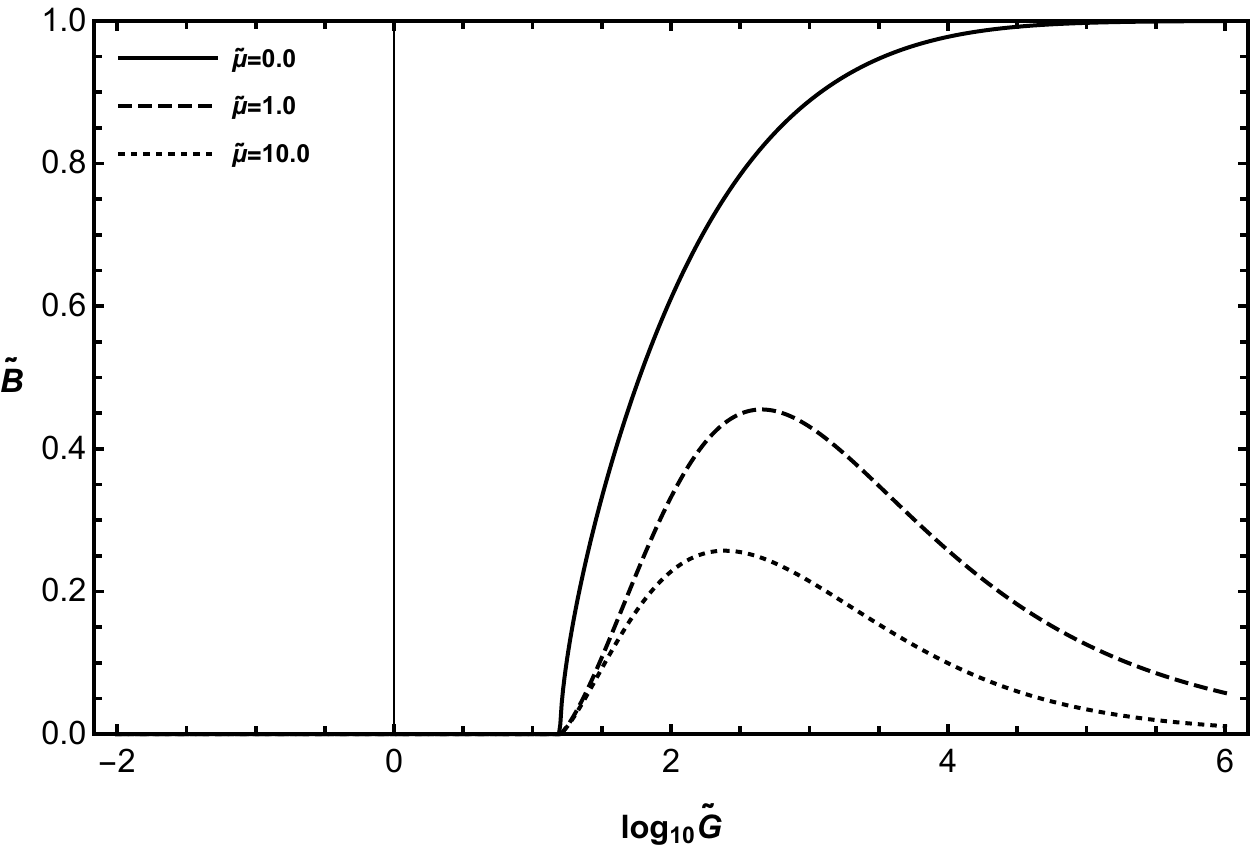} 
\caption{Normalized gap equation solutions in the massless case for: $\mu=0.0$ (black curve),  $\mu=1.0$ (dashed curve), and  $\mu=10.0$, in units of $b$. }
\label{fig5}

\end{figure}

\begin{figure}[htb]
\centering

\includegraphics[width=1.\linewidth]{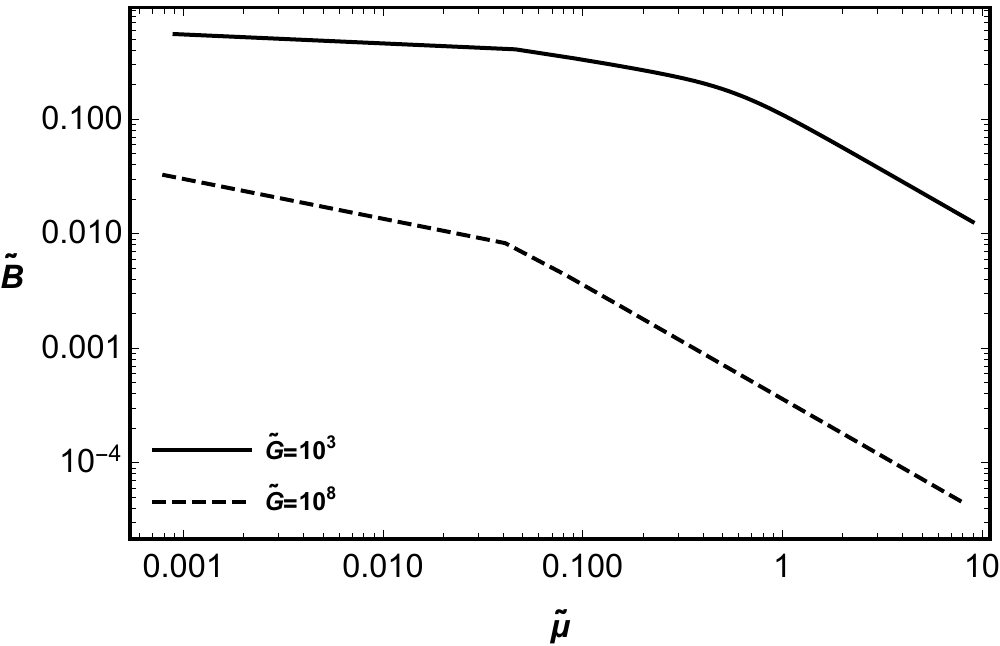} 
\caption{Normalized gap equation solutions in the massless as a function of $\mu$ for: $\tilde{G}=10^3$ (black curve),  $\tilde{G}= 10^8$ (dashed curve), in units of b. }
\label{fig6}
\end{figure}

The solution of the gap equation is given by $B = F(\mu, G)$( with $F(0,G)= b$), can be find numerically and is shown in fig. \eqref{fig5}.  As can be seen in fig. \eqref{fig5} and fig. \eqref{fig6} the effect of the chemical potential is to decrease the value of $B$. 

\subsection{$T \neq 0$ and $\mu \neq 0$:}

\begin{figure}[htb]
\centering
\includegraphics[width=1.\linewidth]{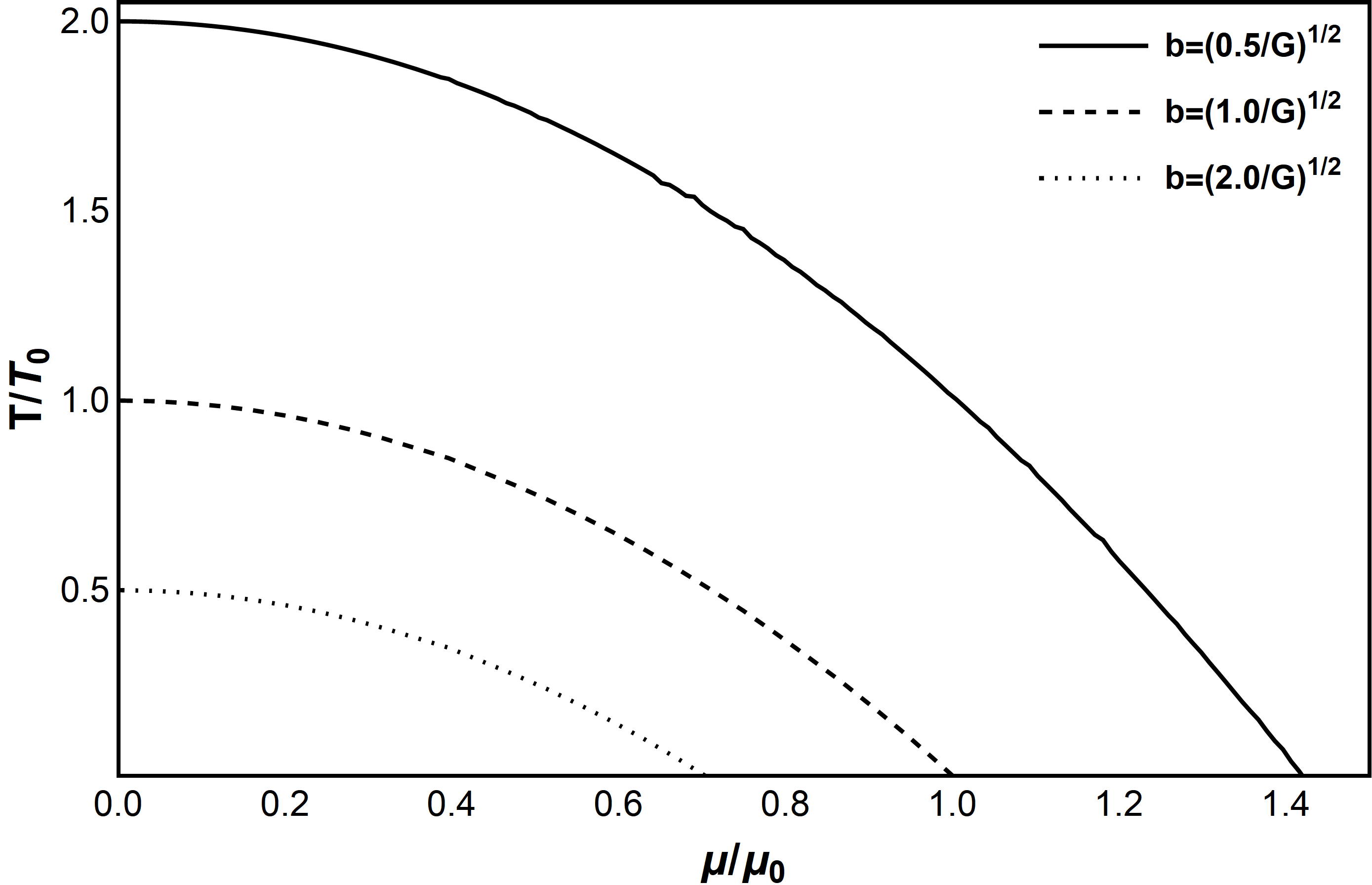} 
\caption{Phase transition diagram for a time-like DLSB with $m=0$; $b= \sqrt{0.5/G}$ (black curve), for $b = \sqrt{G}$ (black and dashed) and for $b = \sqrt{2.0/G}$. Here $T_0 \approx 1.73 G^{-1/2}$ and $\mu_0 \approx 2.55  G^{-1/2}$. }
\label{fig7}
\end{figure}

\begin{figure}[htb]
\centering
\includegraphics[width=1.\linewidth]{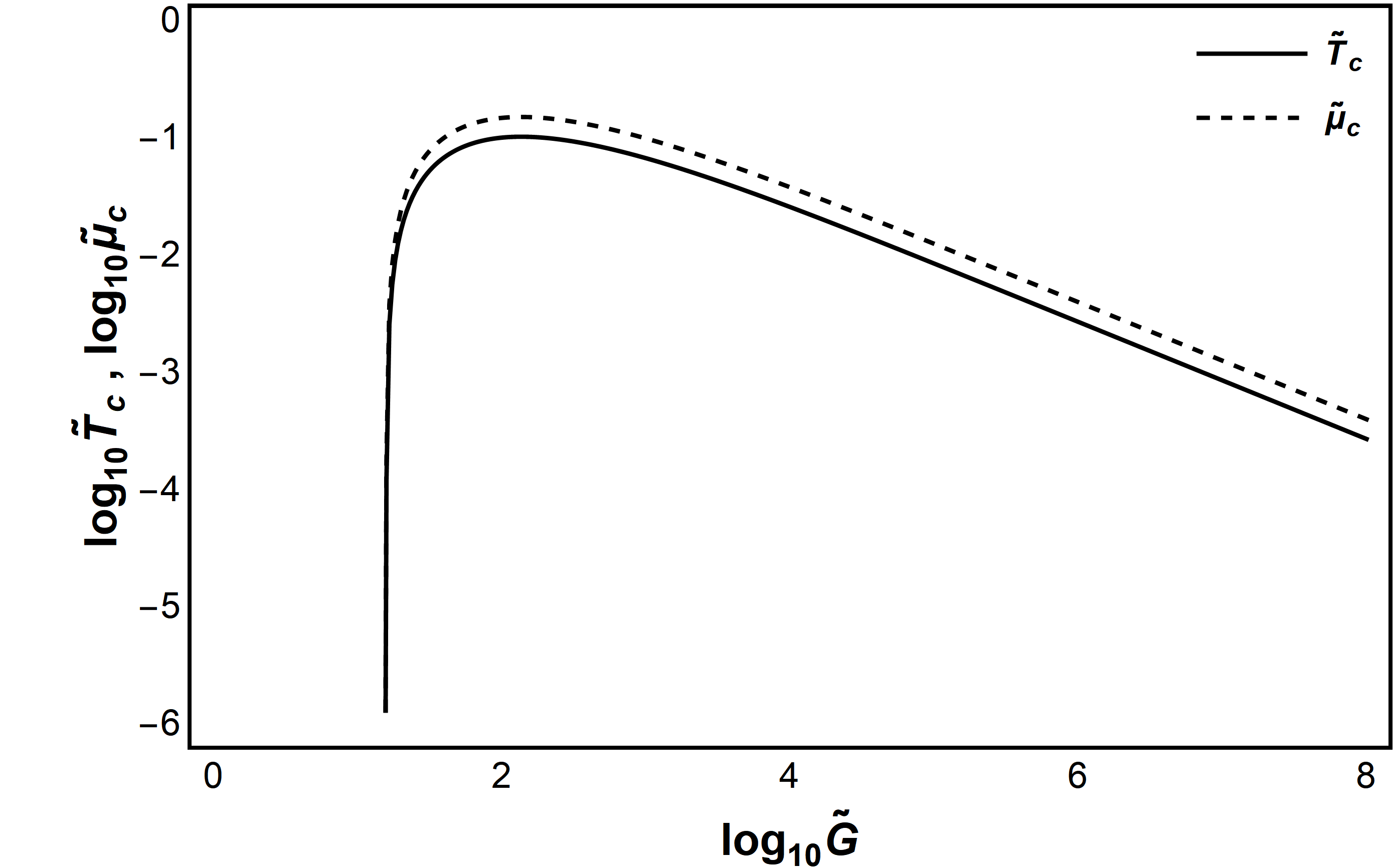} 
\caption{Plot of the $\log_{10}$ of the critical temperature $\tilde{T}_c$ (black curve) and critical chemical potential $\tilde{\mu}_c$  (black and dashed) as a function of $\tilde{G}$ and for $m=0$.}
\label{fig8}
\end{figure}

Finally, from eq. \eqref{gapt1} we can explicitly find the phase diagram through $B(\mu,T,G) = 0$ in the massless case, and the solutions are shown in fig. \eqref{fig7}. We can see that the critical temperature depends upon the magnitude of the DLSB parameter at zero temperature $b$.  Particularly, at zero chemical potential the critical temperature can be written as $T_c \approx  \tilde{b}^2 T_0 $ with $T_0 = 1.73 G^{-1/2}$.  In the zero-temperature limit, the critical chemical potential can be written as $\mu_c \approx \tilde{b}^2 \mu_0 $ with $\mu_0 = 2.55 G^{-1/2}$. Since $b$ is itself a function of $G$ we should combine these expressions with the results of fig. \eqref{fig1} and this combination gives us the fig. \eqref{fig8}. As we can see, the critical temperature and critical chemical potential have the same dependence in $\tilde{G}$, up to a multiplicative constant.  
 
\section{Final comments and perspectives}
In this work, we investigate aspects of dynamical Lorentz symmetry breaking. Firstly, by use of the Keldysh formalism, we find the DS equations for non-equilibrium fermionic systems with interactions given by eq. (1). We also show that through the Keldysh formalism the vacuum expectation value (v.e.v) of the composite field $A_\mu$ given by $g\langle A_\mu \rangle = g\alpha_\mu = 2 B$ will be in general dependent on the non-local coordinate X. This shows that the violation of the Lorentz symmetry and non-equilibrium can be naturally combined by the use of Keldysh formalism.

 Going further, analyzing the DS equations in the equilibrium limit and we see that, through the solution of the DS equations with meaningful approximations, we can access new sectors of the model proposed in \cite{DLSB1,DLSB2}, we reach the solutions of the gap equations for both DLSB and self-energy and we find that system has two special sectors, the first is the weak coupling limit, where the DLSB is dependent of the bare mass $m$. The second sector is the strong coupling limit, where the self-energy vanishes and the DLSB became intense and rules the dynamics of the system (see fig. (2) and fig. (3)). Between these two sectors, we have a transition region where both the DLSB parameter and the self-energy are important to the dynamic of the system.  In fig. \eqref{fig4} we plot the behavior of the DLSB parameter in the small coupling regime ($\tilde{G}<1$) as a function of the mass (assuming the self-energy $\Sigma \approx m$) for $\tilde{G} = 10^{-2}$ and the perturbative solution found in ref.  \cite{DLSB1}. As we can see, the behavior of both curves is similar, and the non-perturbative solution can be seen as an improvement of the perturbative one, without any abrupt change in the character of the DLSB parameter.     
An important difference is that in the limit $m \rightarrow 0$ the DLSB vanishes, whereas the perturbative result is finite ($b\propto G^{-1/2}$).

In particular, we show that in the massless case the DLSB is a possibility of strongly coupled systems with $\tilde{G} > \tilde{G}_c \approx 15.8$. As can be seen in fig. \eqref{fig1} and fig. \eqref{fig2}, there are two different regions concerning the strength of the coupling constant $G$. The first, $\tilde{G}<<\tilde{G}_c$ is characterized by a massless fermionic theory with a null DLSB parameter, and the second, with $\tilde{G}> \tilde{G}_c$ we find a massless fermionic theory and with non-null DLSB parameter.  We show that assuming $m=0$ the DS equations shows that for $\tilde{G}>\tilde{G}_c$, $T_c \approx 1.73 \tilde{b}^2 G^{-1/2}$ at zero chemical potential and $\mu_c \approx 2.55   \tilde{b}^2 G^{-1/2} $ at zero temperature, and the behavior of this critical parameters can be shown in fig. \eqref{fig8}.

 In this context, we can discuss the interpretation of the (time-like) DLSB parameter as a true chiral chemical potential. Due to the existence of a mass in the self-energy gap equation solution, there is no chiral symmetry in the model, unless $m=0$. Therefore, the model will have a true chemical potential only if $m=0$ or in the limit $\tilde{G} \rightarrow \infty$. 
   
In general a chiral chemical potential is related to a conservation of the chiral current, i.e., $\partial_\mu j_5^\mu = 0$, where $j_5^\mu = \bar{\psi}\gamma^\mu \gamma_5 \psi$. Since our model has not a gauge symmetry, there is no anomalies and the symmetry is not broken if $m=0$. Therefore, in the massless case we have a conserved charge given by $Q_5 = \int d^3x j_5^0 = \int d^3 x \left (\psi_L^\dagger \psi_L - \psi_R^\dagger \psi_R\right) = N_L- N_R$, with $L,R$ representing the chiral components of the Dirac fermion. In a thermodynamical perspective, a chiral chemical potential generates a chiral charge density $\rho_5 = \rho_L - \rho_R$ which is called a chiral imbalance and is given by $\rho_5 = \frac{\mu_5^3}{3 \pi^2} + \frac{\mu_5}{3}\left(T^2 + \mu^2 \right)$     \cite{chiraleff}, and in our case $\mu_5 =  g\langle A_0 \rangle  = 2 B$. This effect violates CP and CPT symmetries, such way it can produce a barion number violation and can be a source for Baryogenesis \cite{baryo, baryo2}. 

Even if we assume $m \neq 0$, in the strong coupling limit $\tilde{G}>>\tilde{G}_c$ we can approximate $\Sigma \approx 0$ and in this case, the DLSB parameter is not a true chemical potential and it still contributes to a ``chiral" imbalance, but in this case, we must take into count that the equality between chirality and helicity is not exact. 
From the perspective of Keldysh formalism, a way to implement the chiral chemical potential is to expand the Wigner function $F$ character in Dirac space through projectors $P_{L,R} = \frac{1}{2}(1\pm\gamma_5)$ to accommodate this new feature.   This implementation is beyond the scope of this work and will be investigated in the future.

Going further, a way to improve our results can occur by extending the rainbow approximation through the use of the Bethe-Salpeter equation. We also can improve our result by extending the quenched approximation, taking into account the polarization tensor contribution, and taking into account the non-locality of the renormalization function $A(X,p)$. These modifications will introduce new and more complex non-localities which difficult the extract information, but numerical analysis such as \cite{ref2} can be made.

New features can also appear if we take a more complex structure of the Wigner function $F(X,p)$, particularly the spinorial character. Since the DLSB parameter is proportional to $S^K = 2 i Im [S^R F]$ if we give up the scalar property of the Wigner function a much richer tensor structure arose. In special, $S^K$ could have off-shell contributions which is a feature of non-equilibrium physics and can give us new sources of Lorentz symmetry breaking. Although the model represented by the action \eqref{action1} and analyzed with the non-perturbative tools of the DS equations can be called a toy model, some applications can be thought of in condensed matter, e.g. in the context of 3D Weyl semi-metals models \cite{weyl1}.  The study of these other applications will be the theme of a forthcoming paper.

\begin{acknowledgements}
This work was supported by FAPERJ. The author is grateful to P.C. Malta, P. de Fabritiis, and the referee for interesting discussions and relevant criticism.
\end{acknowledgements}

\section*{Appendix I: Wigner transformation}\label{app1}
Since our results are dependent on the knowledge of the fermionic function, we need to find the solution of the kinetic equation for F(x,y). We can achieve this with the use of the Wigner transformation (WT) which is defined as follows. Let $A(x,x')$ a function and $X = \frac{x+x'}{2}$ and $\tilde{x} = x-x'$, thus:

\begin{equation}
A(X,p) = \int d\tilde{x} e^{- i p \tilde{x}} A\left(X+ \frac{\tilde{x}}{2}, X - \frac{\tilde{x}}{2}\right).
\end{equation}
The inverse of the Wigner transformation is given by:

\begin{equation}
A(x,x') = \int dp e^{i p (x-x')} A\left( \frac{x+x'}{2}, p \right).
\end{equation}

with $dp= \frac{d^dp}{(2 \pi)^d}$ for any dimension $d$. Let $C = A \circ B$ a convolution, the WT of $C$ reads:
\begin{eqnarray}\nonumber
C(X,p) &=& A(X,p) e^{\frac{i}{2}(\overleftarrow{\partial}_X \overrightarrow{\partial}_p- \overleftarrow{\partial}_p\overrightarrow{\partial}_X)} B(X,p) \\\nonumber
&&\hspace{-1.8cm}\approx A(X,p)B(X,p) + \frac{i}{2}\Big( \overrightarrow{\partial}_X A(X,p) \overrightarrow{\partial}_p B(X,p)-\\
&&\hspace{-1.8cm}- \overrightarrow{\partial}_p A(X,p)\overrightarrow{\partial}_X B(X,p) \Big) + ...
\end{eqnarray}
When $A(X,p)$ and $B(X,p)$ is weakly dependent on $X$ coordinate the approximation is valid up to first order in $\partial_X$.  Other important property is the WT of a product of two functions. Let $C(x,x') = A(x,x') B(x,x')$. Thus:
\begin{equation}
C(X,p) = \int dq  A(X,p-q) B(X,q) 
\end{equation}
If the function $A(x,x')$ are invariant under translations, thus $A(x,x') = A(x-x')$ implying $\overrightarrow{\partial}_X A(X,p) = 0$, therefore $A(X,p) = A(p)$ and the WT is equivalent to the Fourier transformation. 

\section*{Appendix II: kinetic equations}\label{app2}

The full fermionic propagator $S_{cd}(x,y)$, as can be reach from eq. \eqref{dysonf}, obey the following equation:
\begin{widetext}
\begin{equation}
\int dy \left[(S_0^{-1})_{cd}(x,y) + g \delta(x-y)\gamma^\mu_{a,cd} \alpha_\mu^a(x) + \Xi_{cd}(x,y) \right]S_{df}(y,z)  = \delta_{cf} \delta(x-z)
\end{equation}
\end{widetext}
Due to the causality structure the full fermionic propagator in the Keldysh formalism it has the following structure in the Keldysh space:
\begin{equation}
S_{df}(y,z) = \begin{pmatrix}S^R(y,z) & S^K(y,z)\\
0 & S^A(y,z) 
\end{pmatrix}
\end{equation}
The self-energy contribution obeys a similar structure, i.e.:
\begin{equation}
\Xi_{df}(y,z) = \begin{pmatrix}\Xi^R(y,z) & \Xi^K(y,z)\\
0 & \Xi^A(y,z) 
\end{pmatrix}
\end{equation}

Therefore the retarded (advanced) component of $S_{df}(y,z)$ obey the following (Kadanoff-baym-like) nonlinear equations:
\begin{eqnarray}\nonumber
&&\left( i \gamma^\mu \partial_\mu + g \gamma^\mu \gamma_5 \alpha^{cl}_\mu(x)- m \right) S^{R(A)}(x,y) +\\ \nonumber
&&+ \int dz \Xi^{R(A)}(x,z) S^{R(A)}(z,y)= \delta(x-y), \\
\end{eqnarray}
where 
\begin{equation}
 g \alpha_\mu^{cl}(x) = i G ~  Tr [\gamma^\mu \gamma_5 S^K(x,x)]
\end{equation}
Important to highlight that the causal structure of the full fermionic propagator fix $\alpha_\mu^q(x) = 0$ unambiguously. Therefore, we use $\alpha_\mu^{cl}(x) = \alpha_\mu(x)$ for short. Since $S^K$ is skew-Hermitian, by parametrization we can write $S^K = S^R \circ F - F \circ S^A$ with $F = F(x,y)$ a generic hermitian tensor in Dirac matrices space and we use the symbol ``$\circ$" as the convolution operator. Therefore, $F$ obey the follow equation:
\begin{eqnarray}\label{kineq2}\nonumber
&&\hspace{-0.5cm}\int dy \left[  \Delta(x,y)F(y,z) - F(x,y) \Delta(y,z) \right] =\Xi^K(x,z) +
\\\nonumber
&&\hspace{0.5cm}+ \int dy \left [ F(x,y)\Xi^R(y,z) - \Xi^A(x,y) F(y,z) \right],\\
\end{eqnarray}
where $\Delta(x,y) = \left( i \gamma^\mu \partial_\mu + g \gamma^\mu \gamma_5 \alpha_\mu(x)- m \right)\delta(x-y)$.   Applying the Wigner transformation in eq. \eqref{kineq2} we find:
\begin{widetext}
\begin{equation}
   Re\left[ \left( \gamma^\mu + \frac{\partial}{\partial p^\mu}\tilde{\Xi}^R(X,p) \right) \frac{\partial}{\partial X_\mu}F(X,p)  -  \frac{\partial}{\partial X^\mu} \tilde{\Xi}^R(X,p)\frac{\partial}{\partial p_\mu}F(X,p) \right] = I_{coll}[F]
\end{equation} 
\end{widetext}
where $\tilde{\Xi}^R(X,p) = \Xi^R(X,p) + g \gamma^\nu\gamma_5 \alpha_\nu(X)$ , $\alpha_\mu(X)$ is the WT of $\alpha_\mu(x)$ and is given by $\alpha_\mu(X) = i  G  ~ \int dp' Tr [\gamma_\mu \gamma_5 S^K(X,p')] $ and $I_{coll}[F] = -i \Xi^K(X,p) + 2 Im[ \Xi^R(X,p) F(X,p)] $ is the collision integral. The equilibrium solution is given by $I_{coll}[F] = 0$ and  can be shown that for fermionic fields in equilibrium $F_{eq}(k) = tanh(\frac{k_0-\mu}{2 T})$ and its connected with the fluctuation – dissipation  theorem (FDT) \cite{kamenevbook}.

\end{document}